\documentstyle[prl,aps]{revtex}
\begin{document}

\def\be{\begin{equation}} 
\def\ee{\end{equation}}

\title{Topological Defects: Fossils from the Early Universe }
\author{Tanmay Vachaspati \\
{\it 
Physics Department,
Case Western Reserve University,
Cleveland OH 44106-7079.
}}

\maketitle

\section*{Abstract}

In the context of current particle physics theories, it is quite 
likely that topological defects may be present in our universe.
An observation of these fossils from the early universe would lead
to invaluable insight into cosmology and particle physics, while
their absence provides important constraints on particle-cosmology
model building. I describe recent efforts to address cosmological 
issues in condensed matter systems such as $^3$He and a possible 
solution to the magnetic monopole problem due to defect interactions.

\section{Introduction}

The electroweak model, GUTs and almost all other particle physics
models are based on ``spontaneous symmetry breaking'' {\it i.e.}
phase transitions. If these descriptions of particle physics are
correct, the inescapable implication is that the early universe 
must have seen phase transitions much like the freezing of water, and, 
the magnetization of iron. Then, the consequences of phase transitions
that we observe in the laboratory can be expected to apply to 
the universe as well. In particular, relics of the high temperature
phase of condensed matter systems called ``topological defects'' are 
routinely observed in the laboratory and similar relics of the 
early high temperature universe could exist in the present universe.
In other words, these are possible fossils from the early universe.
Their observations would be invaluable for gleaning information about
very high energy particle physics and cosmology. It would also have
implications for astrophysics and astronomy. An observed absence of 
topological defects too is very useful since it imposes severe constraints
on particle physics model building. Indeed, the absence of magnetic
monopoles inspired the inflationary revolution in cosmology and GUTs
models are constrained to provide the requisite amount of inflation. 

The hunt for cosmic topological defects depends crucially on their
properties. The last two decades has seen extensive research on 
topological defects and their potential role in cosmology. Very recently,
the lack of experimental input has been relieved by enterprising condensed
matter physicists who have been performing experiments in the laboratory
to answer questions of great interest to cosmologists. 

In this talk I will first describe some recent experimental results obtained 
in $^3$He that have direct implications for cosmology. I will then describe a 
possible new solution to the cosmological monopole over-abundance problem 
highlighting the rich processes involved in the study of topological defects. 

\section{Topological Defects in $^3$He}

Over the last several years, a number of condensed matter
experiments of a cosmological flavour have been performed. 
These are:
\begin{itemize}
\item The experiment in nematic liquid crystals by 
Chuang {\it et. al.} [1990]
where the authors studied the relaxation of a network of strings. 
\item The formation of defects in liquid crystals
by Bowick {\it et. al} [1994] . 
\item The formation of vortices in $^4$He
by Peter McClintock and his group in Lancaster 
[Hendry {\it et. al.}, 1994]. 
\item The ``missing energy'' experiments in $^3$He conducted 
in Grenoble and Lancaster [Bauerle {\it et. al.}, 1996] to study 
the formation of vortices.
\item The ingenious experiment in $^3$He in Helsinki 
[Ruutu {\it et. al.}, 1996] to study the formation and 
distribution of strings. 
\item The $^3$He experiment confirming the analog of the baryon 
number anomaly important for baryogenesis [Bevan {\it et. al.}, 1997]. 
\item The $^3$He experiment studying the conversion of
a baryon number analog into what would correspond to a cosmological
magnetic field [Krusius, Vachaspati and Volovik, 1998] .
\end{itemize}

Here I will summarize the Helsinki experiment studying the distribution
of vortices formed at a phase transition and the violation of the 
$^3$He analog of baryon number.

\subsection{Distribution of Strings}

The first quantity one needs to determine in studying topological defects
is the number density of defects formed during a phase transition. This
was the subject of the McClintock experiment and the Grenoble, Lancaster
and Helsinki experiments. But the Helsinki experiment went further and
determined the distribution of string sizes formed during the phase
transition, {\it i.e.} they obtained the ``spectrum'' of defects. Here
is how they did it.

{}First they prepared a sample of superfluid $^3$He-B at typical temperatures 
of order 1 mK. Then they reheated a bubble in the sample by sending in
neutrons that collided with the $^3$He nucleus and underwent the following
nuclear reaction:
$$
n + {^3}He \rightarrow p + {^3}H_1 + 764 ~ {\rm keV} \ .
$$
The deposited 764 keV restores the symmetry in the bubble (of size about
20 microns) which then cools down and undergoes a phase transition.
(During this phase transition, quasiparticles are produced and their energy
can be detected in the Grenoble and Lancaster experiments. This energy is
short of the injected 764 keV. The missing energy presumably goes into a
network of vortices.) The $^3$He sample in the Helsinki experiment is rotating
which means that there is a relative flow between the superfluid and normal
components of $^3$He in the sample. This relative flow causes a force on the
vortices inside the bubble, that stretches
out some of the vortices (and collapses others). 
The stretched out vortices eventually become straight and settle to
the axis of rotation where they can be counted using NMR.
The spectrum of loop sizes is determined by the Helsinki group since only
loops larger than a critical size can be stretched out by any given rotational
velocity. By varying the rotational velocity, smaller loops can be counted and,
in this way, the spectrum of loop sizes obtained.

The results of the experiment seem to agree with the scale invariant prediction
found by Alex Vilenkin and me in 1984 [Vachaspati and Vilenkin, 1984]. 
Using numerical simulations, we found
that the number density, $dn$, of loops with size between $R$ and $R+dR$ 
is:
$$
dn = \alpha {{dR} \over {R^4}} \ .
$$
The coefficient $\alpha$ depends on the conditions of the phase transition
(for example, the pressure in the $^3$He sample) but the exponent of
$R$ (equal to $-4$) is
the scale-invariant prediction that experiment seems to be confirming.

An important prediction for the cosmology of cosmic strings from our 1984
paper is that the string network consists of infinite strings separate from
the loop population. This has not been seen so far in the experiments. 
However, it is likely that the absence of infinite strings 
may be due to the fixed phase boundary conditions on the surface of the 
bubble. This topic needs further theoretical investigation. 
And then to find the infinite string population,
will need still greater ingenuity on the part of experimentalists.

\subsection{Baryon Number Violation}

In the standard electroweak model, baryon number is violated by anomalous
processes. The classical conservation of the baryonic current is modified to:
\begin{equation}
\partial_\mu j^\mu _{B} = {{N_F} \over {32\pi^2}}
 [ -g^2 W^a _{\mu \nu} {\tilde W}^{a \mu \nu} +
                         {g'}{}^2 Y_{\mu \nu} {\tilde Y}^{\mu \nu} ].
\label{ewanomaly}
\end{equation}
in the usual notation. 
If we consider the case where only the $Z$ gauge field
is non-vanishing and integrate over four volume, this reduces to:
\begin{equation}
\Delta Q_B = N_F {{\alpha ^2 } \over {32\pi^2}} 
cos(2\theta_w ) \Delta \int d^3 x {\vec B}_Z \cdot {\vec Z} 
\label{qb}
\end{equation}
where, $Q_B$ is the baryonic charge and
$\Delta(\cdot )$ represents the difference at two different times. 
The integral on the right-hand side of (\ref{qb}) is the helicity of the 
$Z$ magnetic field
and so, changes in the helicity of electroweak strings 
(which are tubes of $Z$ flux 
for the present purpose) yield changes in baryon number. Now helicity has a
very simple interpretation for $Z$ strings - it is simply the linking number
of two or more loops of string, or, the twist of one string 
[Vachaspati and Field, 1994]. 

This argument made no direct reference to the fermions in the 
standard model. However, the final result can also be 
derived by considering fermion zero modes
on two electroweak string loops that are linked together. 
The fermion zero modes on one of the string 
loops will feel the other loop due to an Aharanov-Bohm interaction. 
This interaction shifts the
Dirac sea and leads to a non-trivial baryon number 
[Garriga and Vachaspati, 1995].
%

In $^3$He, the ingredients of the standard electroweak model are also present
since there are fermionic quasiparticles that play the role of standard model
fermions, and an order parameter that plays 
the role of the gauge and Higgs field.
There is a direct analog of electroweak strings in $^3$He-A and quasiparticle
zero modes too! 
Then the question is if we can observe the analog of baryon number
violation on $^3$He vortices.

It is not possible for me to give all the details in this short report, but
the main points are as follows. In $^3$He, the motion of vortices relative to 
the normal fluid is effectively like an electric field along the vortex.
The quasiparticle zero modes feel this electric field, lifting the
entire Dirac sea along the zero mode. This creates quasiparticles from the
vacuum in direct analogy with the creation of baryon number along electroweak
strings that move across an ambient magnetic field. 

The experiment, however, cannot measure the number of quasiparticles produced 
from the vacuum. Instead it measures the momentum gained by the quasiparticles
which is given by the formula:
$$
\partial_t {\bf P} = {{1} \over {2\pi^2}}
                      \int d^3 x (p_F {\bf {\hat l}}) {\bf E} \cdot {\bf B}
$$
where, $p_F$ is the Fermi momentum and ${\bf {\hat l}}$ is the orientation
of the Cooper pair angular momentum. The way the measurement is made is that
the anomalous change in the quasiparticle momentum leads to an extra force on
the moving vortices. This force is measured, leading to the confirmation
of ``momentogenesis'' along $^3$He vortices and baryogenesis in the standard
model.

The experiment is remarkable confirmation of the physics of anomalies on
vortices which is an important ingredient in 
cosmological baryogenesis scenarios.
(The electroweak sphaleron itself may be viewed as an electroweak 
string [Vachaspati, 1994; Hindmarsh and James, 1994].) The
experiment, however, does not say anything about the creation of matter in the
universe since that depends on the cosmological environment and other factors
such as CP violation and departures from thermal equilibrium.

\section{On the Monopole Problem}

There are three known solutions to the GUT monopole over-abundance
problem. The first is that the GUT phase transition that produces 
magnetic monopoles is followed by a period of inflation that dilutes
the monopole density to acceptable levels [Guth, 1981]. The second is 
that the GUT model includes a period during which electromagnetism is 
broken. During this period, the magnetic monopoles will get connected 
by strings, leading to rapid annihilation and dilution 
[Langacker and Pi, 1980]. The third 
possibility is that the GUT phase transition never occurred and
the universe was always in a state of broken GUT symmetry 
[Dvali, Melfo and Senjanovic, 1995].

All three solutions of the monopole over-abundance problem
require fine tuning of parameters and/or model building solely 
for the purpose of eliminating magnetic monopoles. The solution 
I have recently conjectured in collaboration with Gia Dvali and 
Hong Liu [Dvali, Liu and Vachaspati, 1997]
does not appear to suffer from fine tuning or excessive model
building. The reason I say that the solution is ``conjectured'' is
because it involves a knowledge of defect interactions that is just
beginning to be investigated.

The basic scenario is in the following steps:
\begin{itemize}
\item At the GUT phase transition, magnetic monopoles are formed
together with an infinite unstable domain wall that percolates throughout
space.
\item The domain wall moves through space, sweeping out
the entire volume.
\item When a monopole is hit by the wall, it gets captured. On the
wall the monopole unwinds and its energy spreads and propagates
along the wall. The collision of monopoles and anti-monopoles 
on the wall leads to their annihilation.
\item At some epoch, the domain wall start collapsing as it is
unstable. When the domain wall collapses and disappears, so does all 
the magnetic charge. In this way, neither domain wall nor magnetic
monopole survive today.
\end{itemize}

At every step, there are constraints on the parameters in the
model that need to be satisfied for the scenario to be successful.
In the first step, we need the unstable domain wall to percolate
in space. This imposes a mild constraint on the parameters. In the
second step, we need to make sure that the domain wall sweeps up
nearly all the monopoles in the universe. We know that stable ($Z_2$) 
domain walls will indeed sweep up the whole universe. So the 
constraint here is that the domain
wall cannot be too unstable. The third step assumes that the monopoles
do not pass through the wall, or, if there is a probability associated
with passing through, this is very small. This problem has been studied
in other solitonic systems and by the results available so far, in no
case did the soliton pass through a domain wall [Kudryavtsev, Piette and 
Zakrzewski, 1997; Trebin and Kutka, 1995; Misirpashaev, 1991; Krusius,
Thuneberg and Parts, 1994].  For the fourth step
to be successful, the domain walls should not be too stable, otherwise
they would start dominating the universe. This would impose a constraint
on the model parameters.

In the paper with Dvali and Liu, we evaluated the constraints on the 
parameters and, from our results, it seems that there is no fine tuning
involved. Further, there is no excessive model building involved since
the scenario works in minimal GUT models such as that based on $SU(5)$.
Therefore, the interaction of defects might resolve the monopole problem,
relieving GUTs from the obligation of providing inflation. 

\section{Outlook}

The discovery of cosmic topological defects would provide us with 
a direct window on the early universe. They would yield invaluable
non-perturbative information about particle physics. If
the defects are massive enough, they could be important
astrophysically and may have played a vital role in large-scale 
structure formation.

In this connection, I should mention recent results from 
two groups finding that, in the simplest cosmological model,
GUT scale (gauge) cosmic strings do not 
provide the appropriate anisotropy in the cosmic microwave
background, nor enough power for large-scale structure formation 
[Allen et. al, 1997; Albrecht, Battye and Robinson, 1997]. Recently, 
other cosmological models have also been studied with improvements 
in the fit to observation [Avelino et. al., 1997; Battye, Robinson
and Albrecht, 1997]. This is real progress. But these results
are largely numerical and are hard to test and understand. A clearer 
picture is needed before one can definitively claim that GUT scale 
cosmic strings do not play an astrophysical role. Probably the
cleanest way to rule out, or find, GUT scale cosmic strings is by a
{\it direct} search for them via their gravitational lensing of
background galaxies and quasars [de Laix, Krauss and Vachaspati, 1997].
Results from such searches are expected to be available within the next 
decade.

\

\

\noindent {\it Acknowledgements:} I am grateful to the Department 
of Energy (USA) for research support. 

\vspace{1pc}

\noindent Albrecht, A., Battye, R. A. and Robinson, 
J., Phys. Rev. Lett. {\bf 79}, 4736 (1997).

\noindent Allen, B., Caldwell, R. R., Dodelson, S., Knox, L., 
Shellard, E. P. S.,
and Stebbins, A., Phys. Rev. Lett. {\bf 79}, 2624 (1997).

\noindent Avelino, P. P., Shellard, E. P. S., Wu, J. H. P. and Allen, B.,
astro-ph/9711336.

\noindent Battye, R. A., Robinson, J. and Albrecht, A., astro-ph/9711336.

\noindent Bauerle, C., Bunkov, Yu. M., Fisher, S. N., Godfrin, H.,
and Pickett, G. R., Nature {\bf 382}, 332 (1996).

\noindent Bevan, T. D. C., Manninen, A. J., Cook, J. B., Hook, J. R.,
Hall, H. E., Vachaspati, T. and Volovik, G. E., Nature {\bf 386},
689 (1997).

\noindent Bowick, M. J., Chandar, L., Schiff, E. A. and Srivastava, A. M.,
Science {\bf 263}, 943 (1994).

\noindent Chuang, I., Durrer, R., Turok, N. and Yurke, B., Science {\bf 251},
1336 (1990). 

\noindent de Laix, A. A., Krauss, L. M. and Vachaspati, T., Phys. Rev. Lett.
{\bf 79}, 1968 (1997).

\noindent Dvali, G., Liu, H. and Vachaspati, T., hep-ph/9710301; Phys. Rev.
Lett., in press (1997).

\noindent Dvali, G., Melfo, A. and Senjanovic, G., Phys. Rev. Lett.
{\bf 75}, 4559 (1995). 

\noindent Garriga, J. and Vachaspati, T., Nucl. Phys. {\bf B438},
161 (1995).

\noindent Guth, A. H., Phys. Rev. {\bf D23}, 347 (1981).

\noindent Hendry, P. C., Lawson, N. S., Lee, R. A. M., 
McClintock, P. V. E. and Williams, C. D. H., Nature {\bf 368}, 315 (1994).

\noindent Hindmarsh, M. B. and James, M., Phys. Rev. {\bf D49}, 6109 (1994).

\noindent Krusius, M., Thuneberg, E. V. and 
Parts, U., Physica {\bf B197}, 376 (1994).

\noindent Krusius, M., Vachaspati, T. and Volovik, G. E., in preparation (1998).

\noindent Langacker, P. and Pi, S-Y., Phys. Rev. Lett. {\bf 45}, 1 (1980).

\noindent Kudryavtsev, A., Piette, B. and 
Zakrzewski, W. J., hep-th/9709187 (1997).

\noindent Misirpashaev, T. Sh., Sov. Phys. JETP {\bf 72}, 973 (1991).

\noindent Ruutu, V. M. H., Eltsov, V. B., Gill, A. J., Kibble, T. W. B.,
Krusius, M., Makhlin, Yu. G., Placais, B., Volovik, G. E. and Xu, W., Nature
{\bf 382}, 334 (1996).

\noindent Salomonson, P., Skagertan, B. S. and Stern, A., Phys. Lett. 
{\bf B151}, 243 (1985). 

\noindent Trebin, H-R., and Kutka, R., J. Phys. {\bf A28}, 2005 (1995).

\noindent Vachaspati, T., proceedings of the NATO workshop on ``Electroweak
Physics and the Early Universe'', eds. Romao, J. C. and Freire, F., 
Sintra, Portugal (1994); Series B:
Physics Vol. 338, Plenum Press, New York (1994).

\noindent Vachaspati, T. and Field, G. B., Phys. Rev. Lett. {\bf 73},
373 (1994); Erratum: Phys. Rev. Lett. {\bf 74} (1995).

\noindent Vachaspati, T. and Vilenkin, A., Phys. Rev. {\bf D30}, 2036 (1984).

\noindent Volovik, G. E. and 
Vachaspati, T., Int. J. Mod. Phys. {\bf B10}, 471 (1996).

\vspace{1pc}

\end{document}